\documentclass[a4paper, oneside, twocolumn, notitlepage, 10pt]{extarticle}

\usepackage{paper_style}

\begin{document}
\selectlanguage{english}    
\acrodef{ASIC}[ASIC]{application-specific integrated circuit}
\acrodef{ADC}[ADC]{analog-to-digital converter}

\acrodef{BER}[BER]{bit-error ratio}

\acrodef{CFO}[CFO]{clock frequency offset}

\acrodef{DAC}[DAC]{digital-to-analog converter}
\acrodef{DSP}[DSP]{digital signal processing}
\acrodef{DFB}[DFB]{distributed-feedback}

\acrodef{EB}[EB]{elastic buffer}
\acrodef{EAM}[EAM]{electro-absorption modulator}

\acrodef{FIFO}[FIFO]{first-in first-out}
\acrodef{FPGA}[FPGA]{field-programmable gate array}

\acrodef{LSB}[LSB]{least-significant bit}

\acrodef{MSB}[MSB]{most-significant bit}
\acrodef{MA}[MA]{moving average}

\acrodef{OOK}[OOK]{on-off-keying}

\acrodef{PLL}[PLL]{phase-locked loop}

\acrodef{SNDR}[SNDR]{signal-to-noise-and-distortion ratio}

\acrodef{VCO}[VCO]{voltage-controlled oscillator}


\title{Elastic Buffer Design for Real-Time All-Digital Clock Recovery Enabling Free-Running Receiver Clock with Negative and Positive Clock Frequency Offsets}%


\author{
    Patrick Matalla, Joel Dittmer, Md Salek Mahmud, Christian Koos, and Sebastian Randel}
    
\maketitle                  


\begin{strip}
 \begin{author_descr}

   Institute of Photonics \& Quantum Electronics (IPQ), Karlsruhe Institute of Technology (KIT), Germany.
   
   \centering \textcolor{blue}{\uline{patrick.matalla@kit.edu}}, \textcolor{blue}{\uline{sebastian.randel@kit.edu}}

 \end{author_descr}
\end{strip}

\renewcommand\footnotemark{}
\renewcommand\footnoterule{}


\begin{strip}
    \begin{paper_abstract} 
        We present an elastic buffer design that enables all-digital clock recovery implementation with free-running receiver clock featuring negative and positive clock frequency offsets. Error-free real-time data transmission is demonstrated from $-$\,400\,ppm to $+$\,400\,ppm.  
         \textcopyright2025 The Author(s)
    \end{paper_abstract}
\end{strip}


\section{Introduction}
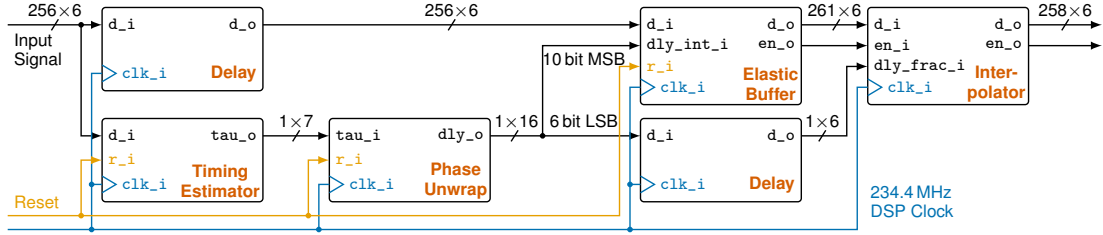
\begin{figure*}[t]
    \centering
    \scalebox{0.92}{\pgfmathsetmacro{\w}{2.3}
\pgfmathsetmacro{\x}{\w/2}
\pgfmathsetmacro{\y}{\w/4}

\begin{tikzpicture}
[line width=0.2mm,
block/.style={draw,fill=white, align=center,rounded corners = 0.05cm, minimum width=\w cm,minimum height=1.2cm},
block2/.style={draw,fill=white, align=center,rounded corners = 0.05cm, minimum width=\w cm,minimum height=1.4cm},
block3/.style={draw,fill=white,align=center,rounded corners = 0.05cm, minimum width=0.6cm,minimum height=0.6cm}]

\scriptsize

\def\strich{
    \draw[] (-0.05,-0.08) -- (0,0) -- (0.05,0.08);
}

\def\clk{
    \draw[draw=pm_blue_dark] (0,-0.12) -- (0.2,0) -- (0,0.12);
    \node[align=left, anchor=west, pm_blue_dark] at (0.19,0) {\texttt{clk\_i}};
}

\def\delay{
    \node[block] at (0,0) {};
    \node[align=right, anchor=east] at (\x,-0.35) {\textbf{\textcolor{pm_red}{Delay}}};
    \node[align=left, anchor=west] at (-\x cm,0.35) {\texttt{d\_i}};
    \node[align=right, anchor=east] at (\x,0.35) {\texttt{d\_o}};
    \begin{scope}[shift={(-\x,-0.35)}] \clk \end{scope}
}

\def\zhu{
    \node[block] at (0,0) {};
    \node[align=center, anchor=east] at (\x+0.05,-0.3) {\textbf{\textcolor{pm_red}{Timing}}\\\textbf{\textcolor{pm_red}{Estimator}}};
    \node[align=left, anchor=west] at (-\x,0.35) {\texttt{d\_i}};
    \node[align=right, anchor=east] at (\x,0.35) {\texttt{tau\_o}};
    \node[align=left, anchor=west] at (-\x,0) {\textcolor{pm_orange}{\texttt{r\_i}}};
    \begin{scope}[shift={(-\x,-0.35)}] \clk \end{scope}
}

\def\pu{
    \node[block] at (0,0) {};
    \node[align=center, anchor=east] at (\x+0.05,-0.3) {\textbf{\textcolor{pm_red}{Phase}}\\\textbf{\textcolor{pm_red}{Unwrap}}};
    \node[align=left, anchor=west] at (-\x,0.35) {\texttt{tau\_i}};
    \node[align=right, anchor=east] at (\x,0.35) {\texttt{dly\_o}};
    \node[align=left, anchor=west] at (-\x,0) {\textcolor{pm_orange}{\texttt{r\_i}}};
    \begin{scope}[shift={(-\x,-0.35)}] \clk \end{scope}
}

\def\eb{
    \node[block2] at (0,0) {};
    \node[align=center, anchor=east] at (\x+0.05,-0.4) {\textbf{\textcolor{pm_red}{Elastic}}\\\textbf{\textcolor{pm_red}{Buffer}}};
    \node[align=left, anchor=west] at (-\x,0.45) {\texttt{d\_i}};
    \node[align=left, anchor=west] at (-\x,0.15) {\texttt{dly\_int\_i}};
    \node[align=left, anchor=west] at (-\x,-0.15) {\textcolor{pm_orange}{\texttt{r\_i}}};
    \begin{scope}[shift={(-\x,-0.45)}] \clk \end{scope}
    \node[align=right, anchor=east] at (\x,0.45) {\texttt{d\_o}};
    \node[align=right, anchor=east] at (\x,0.15) {\texttt{en\_o}};
}

\def\lag{
    \node[block2] at (0,0) {};
    \node[align=center, anchor=east] at (\x+0.05,-0.4) {\textbf{\textcolor{pm_red}{Inter-}}\\\textbf{\textcolor{pm_red}{polator}}};
    \node[align=left, anchor=west] at (-\x,0.45) {\texttt{d\_i}};
    \node[align=left, anchor=west] at (-\x,-0.15) {\texttt{dly\_frac\_i}};
    \node[align=left, anchor=west] at (-\x,0.15) {\texttt{en\_i}};
    \begin{scope}[shift={(-\x,-0.45)}] \clk \end{scope}
    \node[align=right, anchor=east] at (\x,0.45) {\texttt{d\_o}};
    \node[align=right, anchor=east] at (\x,0.15) {\texttt{en\_o}};
}

\node[align=left, anchor=west] at (-2.5,-0.03) {Input\\Signal};
\draw[-latex](-2.5,0.35) -- (-\x,0.35) node[midway, above] {256$\times$6};
\begin{scope}[shift={(-2.5/2-\x/2,0.35)}] \strich \end{scope}
\draw[-latex](-\x-0.3,0.35) -- (-\x-0.3,-1.25) -- (-\x,-1.25);
\draw[fill=black] (-\x-0.3,0.35) circle (0.03);

\begin{scope}[shift={(0,0)}] \delay \end{scope}
\draw[-latex](\x,0.35) -- (-\x+7.7,0.35) node[midway, above] {256$\times$6};
\begin{scope}[shift={(7.7/2,0.35)}] \strich \end{scope}

\begin{scope}[shift={(0,-1.6)}] \zhu \end{scope}
\draw[-latex](\x,-1.25) -- (-\x+3.25,-1.25) node[midway, above] {1$\times$7};
\begin{scope}[shift={(3.25/2-\w/2+\x,-1.25)}] \strich \end{scope}

\begin{scope}[shift={(3.25,-1.6)}] \pu \end{scope}
\draw[] (\x+3.25,-1.25) -- (-\x+7.7-1.4,-1.25) node[midway, above] {1$\times$16};
\begin{scope}[shift={(\x+3.25+0.35,-1.25)}] \strich \end{scope}
\draw[fill=black] (-\x+7.7-1.4,-1.25) circle (0.03);
\draw[-latex] (-\x+7.7-1.4,-1.25) -- (-\x+7.7,-1.25) node[midway, above] {6$\,$bit LSB$\phantom{\times}$};
\draw[-latex] (-\x+7.7-1.4,-1.25) -- (-\x+7.7-1.4,0.05) -- (-\x+7.7,0.05) node[midway, below] {10$\,$bit MSB$\phantom{\times}$};

\begin{scope}[shift={(7.7,-1.6)}] \delay \end{scope}
\draw[-latex](\x+7.7,0.35-1.6) -- (-\x+10.95-0.3,0.35-1.6) -- (-\x+10.95-0.3,-0.25) -- (-\x+10.95,-0.25);
\node[align=center] at (\x+7.7+0.3,0.35-1.6+0.2) {1$\times$6};
\begin{scope}[shift={(\x+7.7+0.3,-1.25)}] \strich \end{scope}

\begin{scope}[shift={(7.7,-0.1)}] \eb \end{scope}
\draw[-latex](\x+7.7,0.35) -- (-\x+10.95,0.35) node[midway, above] {261$\times$6};
\begin{scope}[shift={(\x+7.7+0.5,0.35)}] \strich \end{scope}
\draw[-latex](\x+7.7,0.05) -- (-\x+10.95,0.05);

\begin{scope}[shift={(10.95,-0.1)}] \lag \end{scope}
\draw[-latex](\x+10.95,0.35) -- (2.2+10.95,0.35) node[midway, above] {258$\times$6};
\begin{scope}[shift={(\x+10.95+1.05/2,0.35)}] \strich \end{scope}
\draw[-latex](\x+10.95,0.05) -- (2.2+10.95,0.05);

\node[align=left, anchor=west, pm_blue_dark] at (9.75,-2.2) {234.4\,MHz\\DSP Clock};
\draw[pm_blue_dark](-2.5,-2.6) -- (-\x+10.95-0.15,-2.6) -- (-\x+10.95-0.15,-0.55) -- (-\x+10.95,-0.55) ;
\draw[pm_blue_dark](-\x+7.7-0.15,-2.6) -- (-\x+7.7-0.15,-0.55) -- (-\x+7.7,-0.55) ;
\draw[pm_blue_dark](-\x+7.7-0.15,-1.95) -- (-\x+7.7,-1.95) ;
\draw[fill=pm_blue_dark, draw=pm_blue_dark] (-\x+7.7-0.15,-1.95) circle (0.03);
\draw[fill=pm_blue_dark, draw=pm_blue_dark] (-\x+7.7-0.15,-2.6) circle (0.03);
\draw[pm_blue_dark](-\x+3.25-0.15,-2.6) -- (-\x+3.25-0.15,-1.95) -- (-\x+3.25,-1.95) ;
\draw[fill=pm_blue_dark, draw=pm_blue_dark] (-\x+3.25-0.15,-2.6) circle (0.03);
\draw[pm_blue_dark] (-\x-0.15,-1.95) -- (-\x,-1.95) ;
\draw[pm_blue_dark](-\x-0.15,-2.6) -- (-\x-0.15,-0.35) -- (-\x,-0.35) ;
\draw[fill=pm_blue_dark, draw=pm_blue_dark] (-\x-0.15,-2.6) circle (0.03);
\draw[fill=pm_blue_dark, draw=pm_blue_dark] (-\x-0.15,-1.95) circle (0.03);

\node[align=left, anchor=west, pm_orange] at (-2.5,-2.2) {Reset};
\draw[-latex, pm_orange](-2.5,-2.4) -- (-\x+7.7-0.3,-2.4) -- (-\x+7.7-0.3,-0.25) -- (-\x+7.7,-0.25) ;
\draw[-latex, pm_orange](-\x+3.25-0.3,-2.4) -- (-\x+3.25-0.3,-1.6) -- (-\x+3.25,-1.6) ;
\draw[-latex, pm_orange](-\x-0.3,-2.4) -- (-\x-0.3,-1.6) -- (-\x,-1.6) ;
\draw[fill=pm_orange, draw=pm_orange] (-\x+3.25-0.3,-2.4) circle (0.03);
\draw[fill=pm_orange, draw=pm_orange] (-\x-0.3,-2.4) circle (0.03);

\end{tikzpicture}}	
    \caption{Feedforward clock recovery building blocks for \ac{DSP} hardware implementation.}
    \label{fig:crec_concept}
\end{figure*}

Clock recovery is a fundamental component in communication systems, enabling synchronization of the receiver's sampling clock with the transmitter's in both frequency and phase.
In systems without \ac{DSP}, this is achieved using an analog \ac{PLL} that tunes the receiver \ac{VCO}. Modern high-speed optical transceivers that employ a \ac{DSP}, however, often generate the \ac{VCO} control signal digitally. A timing-error detector and loop filter derive a digital control signal from the sampled received signal, which is then converted into a voltage using a low-speed \ac{DAC} to control the \ac{VCO} \cite{Sun16, fludger2013}. Replacing the analog control path with a fully digital implementation eliminates the need for analog control circuits and the \ac{DAC} and thus fully exploits the advantages of modern CMOS technology, e.g., enhanced power efficiency and reduced chip area \cite{Verbeke18}. An all-digital implementation is a prerequisite for a feedforward clock recovery architecture \cite{matalla_sdm25, matalla21, matalla22}. In such architectures, the estimated sampling phase $\tau$ is digitally corrected via a digital delay element. To do so, $\tau$ is decomposed into an integer sampling period delay $m$ and a fractional sampling period delay $\mu$. Afterwards, an \ac{EB}, which is a type of \ac{FIFO} register with variable read address as well as read/write clocks and bus widths that may differ, compensates for the integer delay $m$ by selecting the appropriate samples. These samples are then interpolated according to $\mu$, typically using a Lagrange interpolator. The overall feedforward clock recovery using the Zhu algorithm \cite{zhu2005} for timing phase estimation is illustrated in Fig.~\ref{fig:crec_concept}.

A key limitation of all-digital clock recovery architectures arises in the presence of \acp{CFO}. In such cases, the \ac{EB} must continuously compensate for the accumulating phase drift between the transmitter and receiver clocks. If the receiver clock is faster than the transmitter clock, the \ac{EB} will be empty at some point (buffer underflow). This can be mitigated by temporarily pausing the \ac{DSP} to allow the buffer to refill. Conversely, if the receiver clock is slower, the \ac{EB} eventually overflows, requiring to drop some samples which results in irreversible data loss. To prevent such scenarios, hybrid clock recovery architectures are commonly employed. These combine a feedforward path for high-frequency jitter compensation with a feedback loop that physically tunes the \ac{VCO} \cite{fludger2013}. One all-digital approach proposed in \cite{schmidt2010} suggests operating the receiver clock marginally faster than the transmitter clock to avoid an \ac{EB} overflow. However, this method presents practical challenges. It requires either separate \acp{VCO} for the transmitter and receiver paths of a transceiver or a shared clock that is slightly up-converted to generate the receiver clock using a \ac{PLL}. Achieving the necessary small frequency offset (within a standardized tolerance of about $\pm$\,20\,ppm, e.g., for a 400ZR standard \cite{400zr}) requires \acp{PLL} with near-unity multiplication factors, which in turn require impractically high divider and multiplier values.

This paper presents an \ac{EB} design that is capable of handling receiver \acp{CFO} lower and higher than the transmitter clock frequency, which facilitates the use of free-running oscillators in transceivers with all-digital clock recovery. The proposed \ac{EB} is implemented on an \ac{FPGA} for a real-time 30-Gbit/s \ac{OOK} optical transmission, demonstrating robust, error-free operation across \acp{CFO} ranging from $-$\,400 to $+$\,400\,ppm.

\vspace{-1em}
\section{Elastic Buffer Concept}
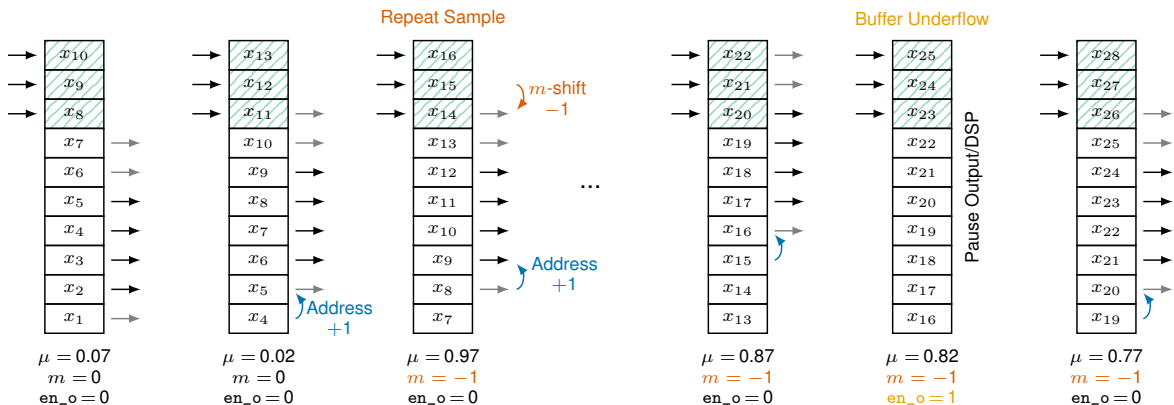
\begin{figure*}[b]
    \centering
    \vspace{-1em}
    \scalebox{0.97}{\begin{tikzpicture}
[line width=0.2mm,
block/.style={draw,align=center,minimum width=0.8cm,minimum height=0.4cm},
]
\scriptsize


\def\datain{
    \draw[-latex](-0.9,0)  -- (-0.5,0);
    \draw[-latex](-0.9,-0.4)  -- (-0.5,-0.4);
    \draw[-latex](-0.9,-0.8)  -- (-0.5,-0.8);
}

\def\dataout{
    \draw[-latex, gray] (0.5,-1.6) -- (0.9,-1.6);
    \draw[-latex, gray] (0.5,-2) -- (0.9,-2);
    \draw[-latex] (0.5,-2.4) -- (0.9,-2.4);
    \draw[-latex] (0.5,-2.8) -- (0.9,-2.8);
    \draw[-latex] (0.5,-3.2) -- (0.9,-3.2);
    \draw[-latex] (0.5,-3.6) -- (0.9,-3.6);
    \draw[-latex, gray] (0.5,-4) -- (0.9,-4);
}

\def\dataoutt{
    \draw[-latex, gray] (0.5,-1.6) -- (0.9,-1.6);
    \draw[-latex, gray] (0.5,-2) -- (0.9,-2);
    \draw[-latex] (0.5,-2.4) -- (0.9,-2.4);
    \draw[-latex] (0.5,-2.8) -- (0.9,-2.8);
    \draw[-latex] (0.5,-3.2) -- (0.9,-3.2);
    \draw[-latex] (0.5,-3.6) -- (0.9,-3.6);
    \draw[-latex, gray] (0.5,-4) -- (0.9,-4);
}

\begin{scope}[shift={(0,0)}]
    \fill[pattern={Lines[angle=45,distance=3pt]}, pattern color=pm_green!40] (-0.4,-0.2) rectangle (0.4,-1.4);

    \node[block]  at (0,-0.4)   {$x_{10}$};
    \node[block]  at (0,-0.8)   {$x_9$};
    \node[block]  at (0,-1.2)   {$x_8$};
    \node[block]  at (0,-1.6)   {$x_7$};
    \node[block]  at (0,-2)     {$x_6$};
    \node[block]  at (0,-2.4)   {$x_5$};
    \node[block]  at (0,-2.8)   {$x_4$};
    \node[block]  at (0,-3.2)   {$x_3$};
    \node[block]  at (0,-3.6)   {$x_2$};
    \node[block]  at (0,-4)     {$x_1$};

    \begin{scope}[shift={(0,-0.4)}] \datain \end{scope}
    \begin{scope}[shift={(0,0)}] \dataout \end{scope}
    
    \node[align=center] at (0,-4.8) {$\mu=\,$0.07\\$m=\,$0\\\texttt{en\_o}\,$=$\,0};
\end{scope}

\begin{scope}[shift={(2.5,0)}]
    \fill[pattern={Lines[angle=45,distance=3pt]}, pattern color=pm_green!40] (-0.4,-0.2) rectangle (0.4,-1.4);

    \node[block]  at (0,-0.4)   {$x_{13}$};
    \node[block]  at (0,-0.8)   {$x_{12}$};
    \node[block]  at (0,-1.2)   {$x_{11}$};
    \node[block]  at (0,-1.6)   {$x_{10}$};
    \node[block]  at (0,-2)     {$x_9$};
    \node[block]  at (0,-2.4)   {$x_8$};
    \node[block]  at (0,-2.8)   {$x_7$};
    \node[block]  at (0,-3.2)   {$x_6$};
    \node[block]  at (0,-3.6)   {$x_5$};
    \node[block]  at (0,-4)     {$x_4$};
    
    \begin{scope}[shift={(0,-0.4)}] \datain \end{scope}
    \begin{scope}[shift={(0,0.4)}] \dataout \end{scope}

    \draw[-latex, pm_blue_dark] (0.5, -4) to[out=20, in=-50] (0.5, -3.65);
    \node[align=center, pm_blue_dark] at (1.1,-4) {Address\\$+$1};

    \node[align=center] at (0,-4.8) {$\mu=\,$0.02\\$m=\,$0\\\texttt{en\_o}\,$=$\,0};
\end{scope}

\begin{scope}[shift={(5,0)}]
    \node[align=center] at (0,0.1) {\textcolor{pm_red}{Repeat Sample}};
    \fill[pattern={Lines[angle=45,distance=3pt]}, pattern color=pm_green!40] (-0.4,-0.2) rectangle (0.4,-1.4);

    \node[block]  at (0,-0.4)   {$x_{16}$};
    \node[block]  at (0,-0.8)   {$x_{15}$};
    \node[block]  at (0,-1.2)   {$x_{14}$};
    \node[block]  at (0,-1.6)   {$x_{13}$};
    \node[block]  at (0,-2)     {$x_{12}$};
    \node[block]  at (0,-2.4)   {$x_{11}$};
    \node[block]  at (0,-2.8)   {$x_{10}$};
    \node[block]  at (0,-3.2)   {$x_9$};
    \node[block]  at (0,-3.6)   {$x_8$};
    \node[block]  at (0,-4)     {$x_7$};
    
    \begin{scope}[shift={(0,-0.4)}] \datain \end{scope}
    \begin{scope}[shift={(0,0.4)}] \dataout \end{scope}

    \draw[-latex, pm_blue_dark] (1, -3.6) to[out=20, in=-50] (1, -3.25);
    \node[align=center, pm_blue_dark] at (1.65,-3.4) {Address\\$+$1};

    \draw[-latex, pm_red] (1, -0.8) to[out=-20, in=50] (1, -1.15);
    \node[align=center, pm_red] at (1.55,-1) {$m$-shift\\$-$1};
    
    \node[align=center] at (0,-4.8) {$\mu=\,$0.97\\\textcolor{pm_red}{$m=-$1}\\\texttt{en\_o}\,$=$\,0};
\end{scope}
    
\node[align=center] at (7,-2.2) {\normalsize...};

\begin{scope}[shift={(9,0)}]
    \fill[pattern={Lines[angle=45,distance=3pt]}, pattern color=pm_green!40] (-0.4,-0.2) rectangle (0.4,-1.4);

    \node[block]  at (0,-0.4)   {$x_{22}$};
    \node[block]  at (0,-0.8)   {$x_{21}$};
    \node[block]  at (0,-1.2)   {$x_{20}$};
    \node[block]  at (0,-1.6)   {$x_{19}$};
    \node[block]  at (0,-2)     {$x_{18}$};
    \node[block]  at (0,-2.4)   {$x_{17}$};
    \node[block]  at (0,-2.8)   {$x_{16}$};
    \node[block]  at (0,-3.2)   {$x_{15}$};
    \node[block]  at (0,-3.6)   {$x_{14}$};
    \node[block]  at (0,-4)     {$x_{13}$};
    
    \begin{scope}[shift={(0,-0.4)}] \datain \end{scope}
    \begin{scope}[shift={(0,1.2)}] \dataout \end{scope}

   \draw[-latex, pm_blue_dark] (0.5, -3.2) to[out=20, in=-50] (0.5, -2.85);

    \node[align=center] at (0,-4.8) {$\mu=\,$0.87\\\textcolor{pm_red}{$m=-$1}\\\texttt{en\_o}\,$=$\,0};
\end{scope}

\begin{scope}[shift={(11.5,0)}]
    \fill[pattern={Lines[angle=45,distance=3pt]}, pattern color=pm_green!40] (-0.4,-0.2) rectangle (0.4,-1.4);

    \node[align=center] at (0,0.1) {\textcolor{pm_orange}{Buffer Underflow}};

    \node[block]  at (0,-0.4)   {$x_{25}$};
    \node[block]  at (0,-0.8)   {$x_{24}$};
    \node[block]  at (0,-1.2)   {$x_{23}$};
    \node[block]  at (0,-1.6)   {$x_{22}$};
    \node[block]  at (0,-2)     {$x_{21}$};
    \node[block]  at (0,-2.4)   {$x_{20}$};
    \node[block]  at (0,-2.8)   {$x_{19}$};
    \node[block]  at (0,-3.2)   {$x_{18}$};
    \node[block]  at (0,-3.6)   {$x_{17}$};
    \node[block]  at (0,-4)     {$x_{16}$};
    
    \begin{scope}[shift={(0,-0.4)}] \datain \end{scope}
    \node[align=center, rotate=90] at (0.7,-2.2) {Pause Output/DSP};

    \node[align=center] at (0,-4.8) {$\mu=\,$0.82\\\textcolor{pm_red}{$m=-$1}\\ \textcolor{pm_orange}{\texttt{en\_o}\,$=$\,1}};
\end{scope}

\begin{scope}[shift={(14,0)}]
    \fill[pattern={Lines[angle=45,distance=3pt]}, pattern color=pm_green!40] (-0.4,-0.2) rectangle (0.4,-1.4);

    \node[block]  at (0,-0.4)   {$x_{28}$};
    \node[block]  at (0,-0.8)   {$x_{27}$};
    \node[block]  at (0,-1.2)   {$x_{26}$};
    \node[block]  at (0,-1.6)   {$x_{25}$};
    \node[block]  at (0,-2)     {$x_{24}$};
    \node[block]  at (0,-2.4)   {$x_{23}$};
    \node[block]  at (0,-2.8)   {$x_{22}$};
    \node[block]  at (0,-3.2)   {$x_{21}$};
    \node[block]  at (0,-3.6)   {$x_{20}$};
    \node[block]  at (0,-4)     {$x_{19}$};
    
    \begin{scope}[shift={(0,-0.4)}] \datain \end{scope}
    \begin{scope}[shift={(0,0.4)}] \dataout \end{scope}

    \draw[-latex, pm_blue_dark] (0.5, -4) to[out=20, in=-50] (0.5, -3.65);

    \node[align=center] at (0,-4.8) {$\mu=\,$0.77\\\textcolor{pm_red}{$m=-$1}\\\texttt{en\_o}\,$=$\,0};
\end{scope}
\end{tikzpicture}}	
    \caption{Elastic buffer concept demonstrating the necessity of repeating a sample and the case of an buffer underflow.}
    \label{fig:eb_concept}
\end{figure*}

A buffer overflow occurs when the receiver samples too slowly. While increasing the receiver sampling rate can prevent this, it is typically avoided due to the reasons mentioned earlier. Digital resampling faces similar limitations: despite efficient polyphase filterbank implementations, the high up- and downsampling ratio quickly leads to large hardware designs in parallel implementation. In this work, we propose to overclock the \ac{EB} without altering the sampling rate, i.e., reading the samples faster from the \ac{EB} than new samples are written. Again, a \ac{PLL} to generate a faster read clock is not suitable due to the conversion ratio constraints and leads to crossing clock domains in hardware. Instead, in \ac{FPGA} or \ac{ASIC} implementations with parallel processing, overclocking is achieved by choosing a read bus width larger than the write bus width. This allows more samples to be read per cycle, with the read address incremented accordingly to prevent duplicate reads. As a result, buffer underflows occur regularly even under negative \acp{CFO}, but buffer overflows are effectively eliminated. 

Fig.~\ref{fig:eb_concept} illustrates the \ac{EB} concept with an input of 3 parallel samples and an output of 4 parallel samples after a Lagrange interpolator. In each clock cycle, input samples are written into the register from the left, while the interpolator reads the required samples from the right. The gray-shaded output samples indicate the additional samples required by the interpolator memory. Since one additional sample is generated per cycle, the read address increments by one each cycle (blue arrows). In case of a phase wrap of the fractional delay, the integer delay has to adjust the address by $\pm\,$1. For negative \acp{CFO}, the integer delay decreases, requiring one sample to be repeated (indicated by a red $m$-shift). Hence, the read address remains unchanged during that cycle (third register). If the read address reaches the buffer end (fourth register), the \ac{EB} is reset in the next cycle and the subsequent \ac{DSP} is paused to avoid double processing. This is signaled by setting the active-low enable signal \texttt{en\_o} high. A key question is how much negative \ac{CFO} this method can tolerate. Assuming a fractional sampling offset is sampled in the interval [0,1) with only two estimates over time, a phase wrap takes place every second cycle, i.e., a sample must be repeated. However, since the read address still advances by one each cycle, even this worst-case scenario avoids buffer overflow. The green-shaded area indicates the samples used for timing estimation. Consequently, the timing offset compensation may incur a maximum latency of one clock cycle, depending on the data read address. However, this offset corresponds to extreme clock jitter or \ac{CFO} and is negligible in practical scenarios.

\vspace{-1em}
\section{FPGA Implementation}
To validate the proposed \ac{EB} concept, we implement the entire feedforward clock recovery architecture on an AMD Virtex UltraScale+ XCVU9P \ac{FPGA}. The hardware modules, including bus widths and port names are depicted in Fig.~\ref{fig:crec_concept}, with bus widths formatted as “number of samples $\times$ bit width per sample”. The system processes 256 parallel input samples per cycle with nominally 2 samples per symbol oversampling. Unless noted otherwise, all data is represented as signed integers. First, the input samples are split into two paths. 
In the lower path, the sampling phase normalized to the sampling rate is estimated using the Zhu algorithm \cite{zhu2005, matalla21, matalla22} with a block size of 256 samples (one clock cycle), followed by a \ac{MA} over 16 cycles for smoothing the timing estimate (as described in \cite{matalla_sdm25}). This produces one sampling offset in the interval [$-$1,1) for a block of 256 samples every cycle, which is then unwrapped in the subsequent module. The phase unwrapping module computes the phase steps, applies a phase unwrap using an edge detector, and accumulates the result over time. The resulting accumulated phase \texttt{dly\_o} has a 16-bit resolution, with the 6 \acp{LSB} (unsigned) representing the fractional component.
The integer delay $m$ and fractional delay $\mu$ can be easily obtained by splitting the phase into the \ac{MSB} and \ac{LSB}, respectively. To align the data path with the estimated sampling offset, the upper path used for sample correction is delayed to match the latency of the estimation path. The \ac{EB} is configured for 256 parallel input samples and 258 output samples after the Lagrange interpolator (to maintain an even output width). Accounting for the Lagrange interpolator memory, a total of 256$\,+\,$2$\,+\,$3$\,=\,$261 samples are read per cycle from the \ac{EB}. The \ac{EB} outputs, along with the appropriately delayed fractional sampling offset, are then processed by a third-order Lagrange interpolator to generate 258 timing-corrected samples. An external reset signal is used to initialize the \ac{MA} filter, phase accumulator, and \ac{EB} address in case of undefined states during \ac{FPGA} startup. Note, that the entire clock recovery system operates within only a single \ac{FPGA} clock domain.

\vspace{-1em}
\section{Real-Time Experiment}
\begin{figure}[htbp]
    \centering
    \input{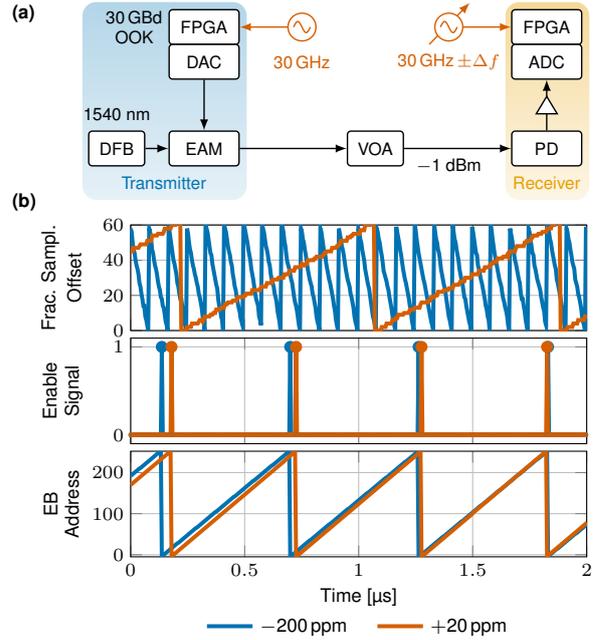}	
    \vspace{-1.6em}
    \caption{(a) Optical back-to-back transmission setup. (b) Fractional sampling offset, enable signal \texttt{en\_0}, and \ac{EB} address for \acp{CFO} of $+$20\,ppm and $-$200\,ppm.}
    \label{fig:setup}
\end{figure}

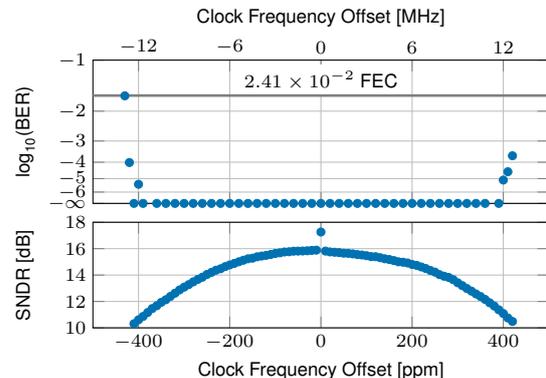
\begin{figure}[htbp]
    \centering
    \scriptsize

\begin{tikzpicture}

\begin{axis}[%
width=6cm,
height=1.9cm,
at={(0cm,0cm)},
scale only axis,
xmin=-500, xmax=500,
ymin=1, ymax=7,
xlabel near ticks,
ylabel near ticks,
y dir=reverse,
ymode=log,
ytick={7,6,5,4,3,2,1},
yticklabels={$-\infty$,$-6$,$-5$,$-4$,$-3$, $-2$, $-1$},
xlabel={Clock Frequency Offset [MHz]},
xticklabel pos=upper,
xtick={-400,-200,0,200,400},
xticklabels={$-12$,$-6$,$0$,$6$,$12$},
ylabel={$\text{log}_{\text{10}}\text{(BER)}$},
axis background/.style={fill=white},
xmajorgrids,
ymajorgrids
]
\node[fill=white, anchor=center, align=center] at (axis cs: 0,1.32) { $2.41\times 10^{-2}$ FEC};

\addplot [color=gray, line width=1pt, forget plot]
  table[row sep=crcr]{%
-500	1.61798295742513\\
500	1.61798295742513\\
};

\addplot [color=pm_blue_dark, only marks, mark size=1.5pt, mark=*, mark options={solid, pm_blue_dark}, forget plot]
  table[row sep=crcr]{%
-500	0.308518893199334\\
-490	0.307428015352968\\
-480	0.309516848723092\\
-470	0.310325386292163\\
-460	0.323017566318015\\
-450	0.336253857582278\\
-440	0.647346768835462\\
-430	1.62429981541203\\
-420	4.01772876696043\\
-410	7\\
-400	5.39794000867204\\
-390	7\\
-360	7\\
-340	7\\
-320	7\\
-300	7\\
-280	7\\
-260	7\\
-240	7\\
-220	7\\
-200	7\\
-180	7\\
-160	7\\
-140	7\\
-120	7\\
-100	7\\
-80	7\\
-60	7\\
-40	7\\
-20	7\\
0	7\\
20	7\\
40	7\\
60	7\\
80	7\\
100	7\\
120	7\\
140	7\\
160	7\\
180	7\\
200	7\\
220	7\\
240	7\\
260	7\\
280	7\\
300	7\\
320	7\\
340	7\\
360	7\\
390	7\\
400	5.09691001300808\\
410	4.55284196865779\\
420	3.66554624884907\\
430	0.531422288733893\\
440	0.403301764018977\\
450	0.342301790065676\\
460	0.317136809952331\\
470	0.308568383030945\\
480	0.307526752305023\\
490	0.307265853362594\\
500	0.307089658979237\\
};
\end{axis}

\begin{axis}[%
width=6cm,
height=1.4cm,
at={(0cm,-1.65cm)},
scale only axis,
xmin=-500, xmax=500,
ymin=10,
ymax=18,
ytick={10,12,14,16,18},
xlabel near ticks,
ylabel near ticks,
ylabel={SNDR [dB]},
ylabel shift=0.24cm,
xlabel={Clock Frequency Offset [ppm]},
axis background/.style={fill=white},
xmajorgrids,
ymajorgrids
]
\addplot [color=pm_blue_dark, only marks, mark size=1.5pt, mark=*, mark options={solid, pm_blue_dark}, forget plot]
  table[row sep=crcr]{%
-500	-2.91872963789621\\
-490	-2.93371452825937\\
-480	-2.90779217600704\\
-470	-2.90288249509001\\
-460	-2.76648851278799\\
-450	-2.61452296354861\\
-440	0.966613700926075\\
-430	7.65167352623405\\
-420	9.97984209885118\\
-410	10.3152426605342\\
-400	10.610592984618\\
-390	10.8662361293854\\
-380	11.1604125143535\\
-370	11.4608514898272\\
-360	11.6768733239648\\
-350	11.9374486676144\\
-340	12.1326378677323\\
-330	12.3845883781323\\
-320	12.626050217182\\
-310	12.874841408734\\
-300	13.083428510926\\
-290	13.2739390062306\\
-280	13.4833015682496\\
-270	13.6707995507422\\
-260	13.8359095048715\\
-250	14.0449062963927\\
-240	14.2048599788965\\
-230	14.3682334528865\\
-220	14.4773212106462\\
-210	14.6411646041344\\
-200	14.7655968211888\\
-190	14.8836445809613\\
-180	15.0077700210068\\
-170	15.1239514968926\\
-160	15.2328711888428\\
-150	15.28125356115\\
-140	15.3837711984252\\
-130	15.4412379145909\\
-120	15.494440865854\\
-110	15.5603630987855\\
-100	15.6377755409177\\
-90	15.6838801788174\\
-80	15.7490899819985\\
-70	15.7672167557917\\
-60	15.8095701387007\\
-50	15.8159243827106\\
-40	15.8229048991248\\
-30	15.8396482482624\\
-20	15.8667190401147\\
-10	15.8879538000499\\
0	17.2608034709568\\
10	15.8174626425738\\
20	15.7530694563795\\
30	15.7296396168799\\
40	15.7007881139613\\
50	15.6564146298751\\
60	15.6439785824101\\
70	15.6034984399204\\
80	15.5519920201402\\
90	15.5204989584204\\
100	15.466848803972\\
110	15.4008661594324\\
120	15.3913299969804\\
130	15.3142842572732\\
140	15.2340184746771\\
150	15.1775521215195\\
160	15.0964966552225\\
170	15.0043174735963\\
180	14.9753727843634\\
190	14.9184978828093\\
200	14.8038705812435\\
210	14.7344110168214\\
220	14.6162522424697\\
230	14.4966843529601\\
240	14.3497663743172\\
250	14.2249945312772\\
260	14.0355033716039\\
270	13.9131731401714\\
280	13.8242033277928\\
290	13.6541599981658\\
300	13.4165834036152\\
310	13.1994443491132\\
320	12.9823828568332\\
330	12.7851634814539\\
340	12.5697336282947\\
350	12.3756605170294\\
360	12.1552660843014\\
370	11.9024568804361\\
380	11.6556853643524\\
390	11.3689336110075\\
400	11.0879080156616\\
410	10.7489896941288\\
420	10.4847419729207\\
430	-0.826876075700284\\
440	-2.03804880803756\\
450	-2.54964264400968\\
460	-2.82994272185169\\
470	-2.92241286634288\\
480	-2.92283002472641\\
490	-2.93730219158761\\
500	-2.93701376858155\\
};
\end{axis}

\end{tikzpicture}%
    \vspace{-0.3em}
    \caption{Experimental results of the \ac{BER} and \ac{SNDR} over \ac{CFO} in units MHz and ppm obtained from $2^{18}$ received symbols.}
    \label{fig:results}
\end{figure}

The real-time clock recovery is validated in an IM/DD optical back-to-back system with 30-Gbit/s \ac{OOK} as depicted in Fig.~\ref{fig:setup}(a). The transmitter and receiver processing units each consists of a Keysight USPA platform equipped with an AMD \ac{FPGA}. The transmitter generates a 30\,Gbit/s NRZ signal, which modulates a 1540-nm \ac{DFB} laser using a 34-dB 3-dB-bandwidth \ac{EAM} (Optilab LT-40-E-M). The optical power is attenuated to $-$1\,dBm to fully utilize the \ac{ADC} dynamic range after detection by a 27-GHz 3-dB bandwidth photodiode (Optilab PR-40G-M). The \ac{ADC} samples the signal at 60\,GSa/s, resulting in nominally twofold oversampling as required by the clock recovery. The clock recovery is performed in real-time on the \ac{FPGA} and afterwards $2^{18}$ received symbols, along with the fractional sampling offset, \ac{EB} address, and \texttt{en\_o} signal, are written into a memory for analysis. To assess performance under various \acp{CFO}, the receiver clock is detuned by manually changing the external oscillator frequency. Fig.~\ref{fig:setup}(b) illustrates the timing behavior for \acp{CFO} of $+$20\,ppm and $-$200\,ppm. As expected, a higher \ac{CFO} results in more $+$1 address corrections and thus less closely spaced \texttt{en\_o} signals.
Fig.~\ref{fig:results} presents the \ac{SNDR} (without any equalizer) and \ac{BER} derived from the recorded data as a function of \ac{CFO}. As the timing estimate update rate is 60\,GHz/256 and the 16-tap \ac{MA} filter has a 3-dB bandwidth of about 1/32-th and the first spectral null at about 1/16-th of the update rate, the 
3-dB clock recovery bandwidth results to 7.32\,MHz and a failure to track the \ac{CFO} is expected at 14.65\,MHz. These values align well with the observed clock recovery failure in Fig.~\ref{fig:results}. The \ac{SNDR} at 0\,ppm is 1.5\,dB higher than at adjacent \acp{CFO}. This \ac{CFO}-related effect arises from a dynamic common-mode voltage control of the \ac{ADC}, which alters the signal amplitude for different sampling offsets due to the changing sample statistics. Advanced \ac{ADC} designs in transceivers can mitigate this behavior.

\vspace{-0.5em}
\section{Conclusions}
We present an \ac{EB} design that enables all-digital clock recovery with a free-running receiver oscillator, supporting, both, positive and negative \acp{CFO}. The clock recovery is implemented on an \ac{FPGA}, demonstrating error-free data transmission for \acp{CFO} up to $\pm\,$400\,ppm. This method eliminates the need for analog \ac{VCO} control and a low-speed \ac{DAC}, offering an important step toward fully digital, power-efficient clock recovery in modern \ac{DSP}-based optical transceivers.
\vspace{-1em}

\clearpage
\section{Acknowledgements}
This work was supported by the German Federal Ministry of Education and Research (German: Bundesministerium f\"ur Bildung und Forschung) in the Open6GHub project (Grant 16KISK010) and the HYPERCORE project (Grant 16KIS2103). 

\printbibliography

\end{document}